\documentclass[reprint,
 amsmath,amssymb,
 aps,
 prd,
]{revtex4-1}

\usepackage{graphicx}
\graphicspath{{img/}}
\usepackage{dcolumn}
\usepackage{bm}
\usepackage{hyperref}
\usepackage{cleveref}
\usepackage[separate-uncertainty=true]{siunitx}

\usepackage[utf8]{inputenc}
\usepackage[T1]{fontenc}

\newcommand{\componentlibrary}{{This sketch was created using the \href{http://www.gwoptics.org/ComponentLibrary/}{Component Library by Alexander Franzen}, licensed under \href{https://creativecommons.org/licenses/by-nc/3.0/}{CC BY-NC 3.0}.}}

\begin{document}

\title{Dual balanced readout for scattered light noise mitigation in Michelson interferometers}

%%% AUTHOR INFORMATION
\author{André Lohde}
 \email{a.lohde@vu.nl}
 \altaffiliation{Author moved to Vrije Universiteit Amsterdam.}
\author{Daniel Voigt}
 \email{daniel.voigt@physik.uni-hamburg.de}
\author{Oliver Gerberding}
 \email{oliver.gerberding@physik.uni-hamburg.de}

\affiliation{Institute of Experimental Physics,
University of Hamburg,
Luruper Chaussee 149,
22761 Hamburg,
Germany}

\date{\today}

\begin{abstract}
    Ground-based gravitational wave detectors use laser interferometry to detect the minuscule distance change between test masses caused by gravitational waves. Stray light that scatters back into the interferometer causes transient signals that can cover the same frequency range as a potential gravitational wave signal. This scattered light noise is a potentially limiting factor in current and future detectors thus making it relevant to find new ways to mitigate it. Here, we demonstrate experimentally a technique for the subtraction of scattered light noise from the displacement readout of a Michelson interferometer. It is based on using a balanced homodyne detector at both the symmetric and the antisymmetric port. While we have been able to demonstrate a noise reduction of \SI{13.2}{\decibel}, the readout scheme seems to be only limited by the associated noise couplings, i.e. shot noise and the coupling of laser noise. We also discuss challenges for using the dual balanced homodyne detection scheme in more complex interferometer topologies, which could lead to improvements in scattered light noise mitigation of gravitational wave detectors.
\end{abstract}

\maketitle

\section{Introduction}
    Ground-based gravitational wave detectors, such as Advanced LIGO \cite{LIGOScientific}, Advanced Virgo \cite{VIRGO}, KAGRA \cite{KAGRA}, GEO600 \cite{Luck:2010} and the future Einstein Telescope (ET) \cite{ET} and Cosmic Explorer (CE) \cite{CosmicExplorer, CosmicExplorer2} employ ultra-precise displacement measurements between free floating test masses to detect gravitational wave signals in the frequency band of \SI{3}{\hertz} to \SI{10}{\kilo\hertz}. Michelson laser interferometers, enhanced using optical resonators in the long arms and for power and signal recycling, are the central topology for these detectors.
    Their sensitivity is limited by various types of noise. Among those is scattered light noise, created by light scattered out of the interferometer and back into the readout with an additional, parasitic phase dynamic \cite{billing1983,vinet1996,vinet1997a}. Its non-linear coupling up-converts low-frequency signals into the measurement band \cite{schilling_method_1981,schnupp_reduction_1985} and e.g. creates transient signals, called scatter glitches. Scattered light couplings are a limiting factor at low frequencies \cite{martynov_2016,Canuel2013} and might contribute to yet unidentified noise. Various measures are currently implemented in operation to reduce the scattered light noise, including the heavy use of baffles and the reduction of scattering surface motion relative to the test masses \cite{Soni2021,Nguyen2021,Soni2024}. Yet, reaching the design sensitivity of future detectors remains challenging, as even single scattering photons are detrimental to detector performance \cite{Ottaway2012-vh, Chua_2014}.

    In this letter, we study a scheme using two balanced homodyne detectors to read out orthogonal quadratures at the antisymmetric and the symmetric interferometer port (i.e. the commonly used output port and the back-reflection towards the laser) of a Michelson interferometer, allowing for the subtraction of scattered light noise. 

    In earlier works, \textcite{steinlechner_quantum-dense_2013, Meinders2015, ast2016, ast_quantum-dense_2017} describe a readout of both amplitude and phase quadrature at the antisymmetric port of a Michelson interferometer in combination with two-mode squeezing, to achieve what they call quantum-dense metrology. They show that by reading out the phase quadrature in addition to the amplitude quadrature, it is possible to distinguish a signal created by scattered light noise from a length variation of the interferometer arms - which represents a gravitational wave signal. Initially frequencies, at which scattering occurs, were simply proposed to be vetoed out in the detector output data stream, however the technique was developed further to model scattered light noise in the amplitude quadrature from information obtained by measurement of the phase quadrature. This allowed to subtract the modeled scattered light signal and suppress the noise below shot-noise level to recover their injected “gravitational wave”-signal.

    \textcite{Fleddermann2018, steier2008a} showed that back-scattered light can be strongly suppressed in heterodyne laser interferometers by using balanced detection of both output ports. This was studied and demonstrated in the scope of studying the Laser Interferometer Space Antenna (LISA) space-based gravitational wave detector. Specifically, back-scattered light stemming from a fiber directly linking the two optical benches in a LISA satellite can be strongly suppressed, which is necessary to meet LISA mission requirements. In their case the suppression was demonstrated by a factor of up to 60 in amplitude, translating into about \SI{36}{\decibel} of scattered light noise suppression in the readout of the non-reciprocal phase noise - i.e. the measurement of phase noise stemming from a fiber that is not common to both propagation directions. 

    In this paper we study a scheme that is, to some degree, a combination of the two described techniques. In our scheme we locate one balanced homodyne detector at the symmetric port, i.e. reading out the back-reflection from the central beam splitter, while the other one is again placed at the antisymmetric port.

    The obtained signals are combined to read out both interferometer arms individually and thus effectively remove scattered light noise under certain conditions. In the following, we present our findings from investigating the case of a simple Michelson interferometer. First, a theoretical treatment of the signals read out with balanced homodyne detectors is given in \cref{sec:theory}. An experimental demonstration of scattered light removal is described in \cref{sec:experiment}. Finally, in \cref{sec:advanced}, a study of the potential application of the scheme in more complex Michelson interferometer topologies is given.

\section{Theory}
\label{sec:theory}
    An electric field oscillating at frequency $\omega$ with the complex amplitude $\alpha(t)$ can be defined as
    \begin{equation}
        E(t) = \frac{1}{2}\alpha(t)\exp(i\omega t) + c.c.
    \end{equation}
    Then one quadrature $Q(t)$ is defined to be the real part of the field amplitude $\alpha(t)$ and the other quadrature $P(t)$ is defined to be the imaginary part. For the treatment of scattered light noise a laser field is assumed, described by the field amplitude
    \begin{equation}
        \alpha_0(t) = E_0
    \end{equation}
    entering the interferometer from the symmetric port. The field picks up additional phases as it propagates through the interferometer: $\Phi_{X/Y} = 2 k L_{X/Y}$ is the phase picked up due to propagation from the central beam splitter to the X/Y end mirror and back, where $k$ is the wavenumber and $L_{X/Y}$ the macroscopic arm length of the X/Y interferometer arm respectively.
    It is considered that parasitic fields enter both interferometer arms with amplitudes $|\alpha_{sc}| \ll |\alpha_0|$, treated as additional terms appearing in the resulting electric fields at symmetric and antisymmetric port.
    One obtains the following expressions for the quadratures at the antisymmetric interferometer port:
    \begin{equation}
        \begin{split}
        Q_{as} (t) =& - \frac{E_0}{2} \left[\sin(\Phi_Y) + \sin(\Phi_X)\right]  \\
                    & + \frac{1}{\sqrt{2}}\left[ - E_{sc,Y} \sin(\phi_{sc,Y}(t)) + E_{sc,X} \cos(\phi_{sc,X}(t))\right]\\ 
        P_{as} (t) =& \frac{E_0}{2} \left[\cos(\Phi_Y) + \cos(\Phi_X)\right] \\
                    & + \frac{1}{\sqrt{2}}\left[E_{sc,Y} \cos(\phi_{sc,Y}(t)) + E_{sc,X} \sin(\phi_{sc,X}(t))\right]
        \end{split}
        \label{eq:q_as_port}
    \end{equation}
    with $\phi_{sc,X/Y}$ being the phase of the parasitic field originating in the east/north interferometer arm. At the symmetric interferometer port, the quadratures can be expressed as:
    \begin{equation}
        \begin{split}
            Q_{s} (t) =& \frac{E_0}{2}\left[\cos(\Phi_Y) - \cos(\Phi_X)\right]  \\
                       & + \frac{1}{\sqrt{2}}\left[E_{sc,Y} \cos(\phi_{sc,Y}(t)) - E_{sc,X} \sin(\phi_{sc,X}(t))\right] \\
            P_{s} (t) =& \frac{E_0}{2}\left[\sin(\Phi_Y) - \sin(\Phi_X)\right] \\
                       & + \frac{1}{\sqrt{2}}\left[E_{sc,Y} \sin(\phi_{sc,Y}(t)) + E_{sc,X} \cos(\phi_{sc,X}(t))\right].
        \end{split}
        \label{eq:q_s_port}
    \end{equation}
    It can be observed that different combinations of the two quadratures allow for the elimination of contributions from the X or the Y interferometer arm. Assuming a readout of the $Q$ quadrature at the antisymmetric port and the $P$ quadrature at the symmetric port, following combinations arise:
    \begin{equation}
        \begin{split}
            S_1(t) =& \frac{1}{2} \left( Q_{as} (t) + P_s (t) \right) \\
                   =& - \frac{1}{2} E_0 \sin(\Phi_X) + \frac{1}{\sqrt{2}} E_{sc,X} \cos(\phi_{sc,X}(t)) \\
            S_2(t) =& \frac{1}{2} \left( P_s (t) - Q_{as} (t) \right) \\
                   =& \frac{1}{2} E_0 \sin(\Phi_Y) + \frac{1}{\sqrt{2}} E_{sc,Y} \sin(\phi_{sc,Y}(t)).
        \end{split}
        \label{eq:combined_signals_1}
    \end{equation}
    These simple arithmetic combinations show that the field contributions from each arm can be read out individually. When also considering the readout of the respective orthogonal quadrature at both ports, it is possible to obtain two more signals:
    \begin{equation}
        \begin{split}
            S_3(t) =& \frac{1}{2} \left( Q_s (t) + P_{as} (t) \right) \\
                   =& \frac{1}{2} E_0 \cos(\Phi_Y) + \frac{1}{\sqrt{2}} E_{sc,Y} \cos(\phi_{sc,Y}(t)) \\
            S_4(t) =& \frac{1}{2} \left( P_{as} (t) - Q_s (t) \right) \\
                   =& \frac{1}{2} E_0 \cos(\Phi_X) + \frac{1}{\sqrt{2}} E_{sc,X} \sin(\phi_{sc,X}(t)).
        \end{split}
        \label{eq:combined_signals_2}
    \end{equation}
    In each of the four signals defined here, only contributions from either the X or the Y interferometer arm are present, allowing for an individual arm readout. They further allow for scattered light noise mitigation, e.g. when scattering occurs only in one interferometer arm or the contribution of scattering is unequal in the interferometer arms. Thus, a readout of orthogonal quadratures at the symmetric and the antisymmetric port offers the possibility of mitigating scattered light noise. 

    Notably, the choice of an operating point for the interferometer does not have a direct impact on the signal combinations when not considering shot noise and other readout effects and does not seem to impact the dual balanced readout.

\section{Experimental realization}
\label{sec:experiment}
    The readout of two orthogonal quadratures can be experimentally realized by implementing two balanced homodyne detectors, as these are sensitive to a quadrature selected by tuning the phase of the local oscillator. The measured signal is the result of interference between a local oscillator beam serving as a phase reference with the signal to be probed and is proportional to \citep[p. 141]{Fox2006-yd}:
    \begin{equation}
        \text{BHD}(t) \propto E_{lo} \left( \sin(\Psi_{lo}) Q_{sig} - \cos(\Psi_{lo}) P_{sig}\right).
    \end{equation}
    Here, we show a proof-of-principle of the scattered light noise mitigation scheme. We set up a Michelson interferometer and intentionally introduced a deterministic scattered light signal to simulate noise, which we were able to partially subtract from the readout of the antisymmetric port, using the additional information from the readout of the symmetric port.

    Next, the setup is described in \cref{sec:exp_setup} and the results are shown and discussed in \cref{sec:exp_results} and \cref{sec:exp_discussion} respectively.

    \subsection{Setup}
    \label{sec:exp_setup}
        \begin{figure}[!t]
            \centering
            \includegraphics[width=0.4\textwidth]{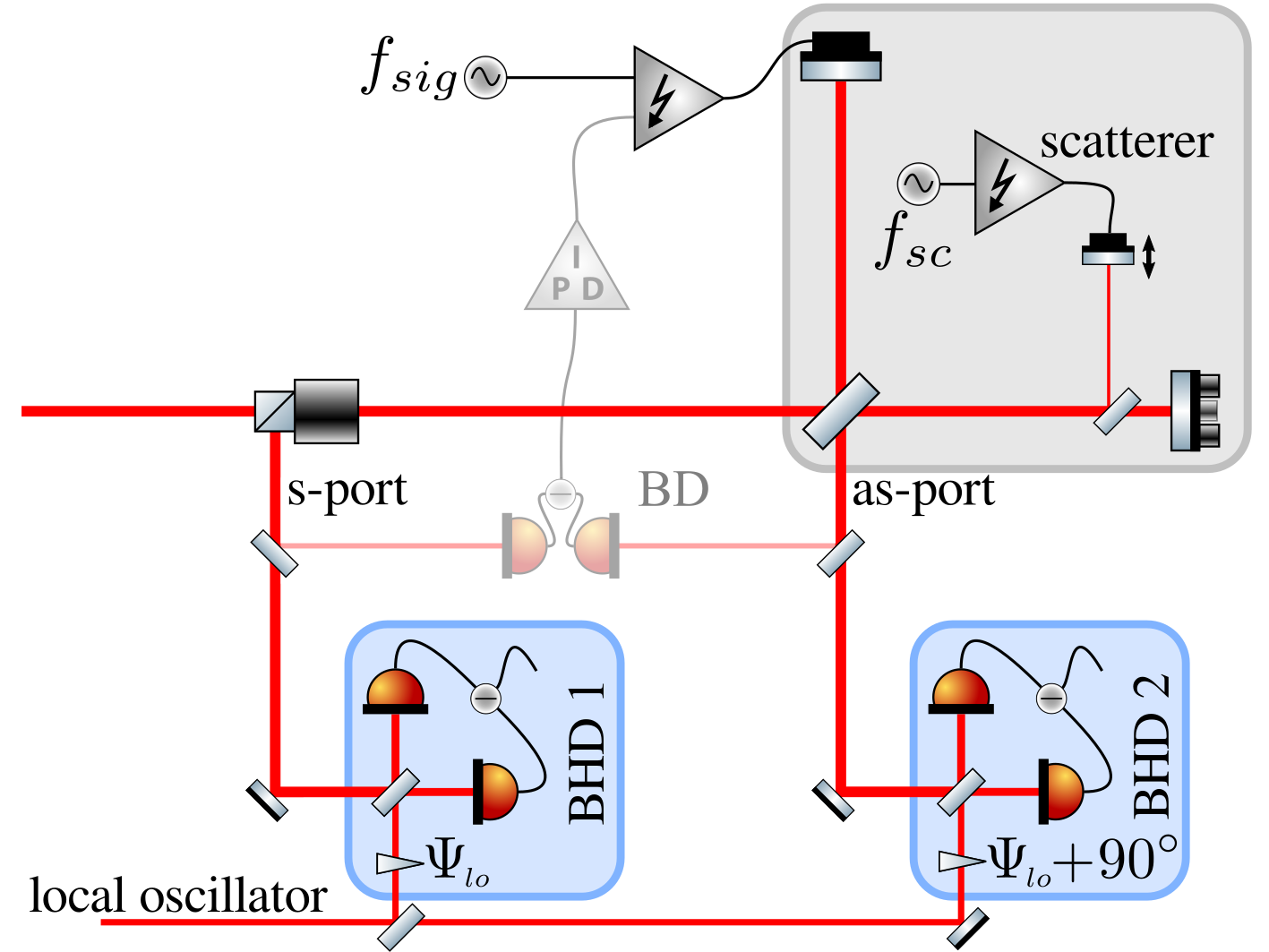}
            \caption{A sketch of the experimental setup for the test of the dual balanced readout. The Michelson interferometer featured a piezoelectric-actuated (PZT) end mirror in one of the arms – referred to as the Y arm – while in the other arm (X arm), some light was scattered out of the beam using a low-reflectivity window and back-scattered from another PZT. The interferometer is operated at mid fringe with a simple locking scheme for which the balanced detector (BD) is used. The symmetric port balanced homodyne detector (BHD1) reads out the $P$ quadrature, while the antisymmetric port balanced homodyne detector (BHD2) reads out the $Q$ quadrature. \componentlibrary}
            \label{fig:experiment}
        \end{figure}

        The setup, consisting of a Michelson interferometer with a total of three readouts, is depicted in \cref{fig:experiment}. The interferometer was kept at mid fringe by using the differential power between the two interferometer ports as the error signal for the actuation of the Y interferometer arm, detected by the balanced readout (BD). Two balanced homodyne detectors were used to read out the interferometer at the antisymmetric and at the symmetric port. The local oscillator beam was picked up in front of the Faraday rotator (FR) used to direct the back-reflected light towards the symmetric port readout. Its phase was tuned for a readout of the quadrature $P_s$ at the balanced homodyne detector of the symmetric port (BHD1) and for a readout of the quadrature $Q_{as}$ at the antisymmetric port (BHD2).

        Scattered light was introduced into the interferometer using a partially reflective ``scatter window'' (reflectance of a few percent) in the X-arm of the interferometer. The diverted stray light beam was directed towards a piezo-mirror -- actuated for a modulation at scatter frequency $f_{sc}$ -- and then back-reflected into the interferometer. This lead to a stray light amplitude of a few percent of the carrier beam. In addition to this scattered light signal, a modulation of the Y-arm piezo-mirror at signal frequency $f_{sig}$ was introduced. These modulations were kept at frequencies far above the control loop bandwidth, such that the interferometer was essentially "free running" and the signals could be observed directly by reading out the detector signal rather than the control signal. It was further intended to find a virtually flat noise floor, which was given at the chosen frequency around \SI{1.2}{\mega\hertz}. Consequently, the motional amplitude of the scatter-mirror was limited at this frequency by the capacity of its piezo and the slew rate of the power supply. The amplitude was further strictly smaller than one wavelength (\SI{1064}{\nano\metre}), as no harmonics were observed, which would have been the case for larger motional amplitudes. The piezos were driven using custom-made voltage amplifier and PID-controller, and a standard signal generator. To achieve sufficient modulation depth at these high frequencies we modulated the piezos at resonance frequencies.

    \subsection{Results}
    \label{sec:exp_results}
        Short time series of \SI{20}{\milli\second} were recorded, acquiring the signals from all three detectors. The calibration factor of the balanced detector BD was determined to be 
        \SI{31.2(0.2)}{\nano\metre\per\volt} with a calibration measurement by ramping through several fringes using the piezo-electric actuable mirror in the interferometer's Y arm. Using this calibration factor the BD signal was calibrated and both BHD signals were normalized to contain the same signal power at \SI{1.24}{\mega\hertz}, which is the signal injected in the Y arm. The following figure \cref{fig:bhd_signals} shows the amplitude spectral densities (ASD) for the displacement signal in all three channels around $f_{sig}$ and $f_{sc}$. To compute the ASD's, Welch's method \citep{Welch1967} was used with a bandwidth of \SI{50}{\hertz}, an overlap of \SI{50}{\percent} and Blackman windowing. It shows that all three detector signals detect some scattered light signal, although at different levels. Assuming that BHD1 truly measures $P_s$ and BHD2 measures $Q_{as}$, it would rather be expected that both signals inhibit the same scattered light power, since the signal should show in the same way, according to \cref{eq:q_s_port,eq:q_as_port}. Furthermore both BHD readouts inhibit a higher overall readout noise level at around \SI{60}{\femto\meter\per\sqrt{\hertz}} in contrast to a noise level of about \SI{12}{\femto\meter\per\sqrt{\hertz}} in the BD signal.

\begin{figure*}[!t] % spanning the whole page
            \includegraphics[width=0.9\textwidth]{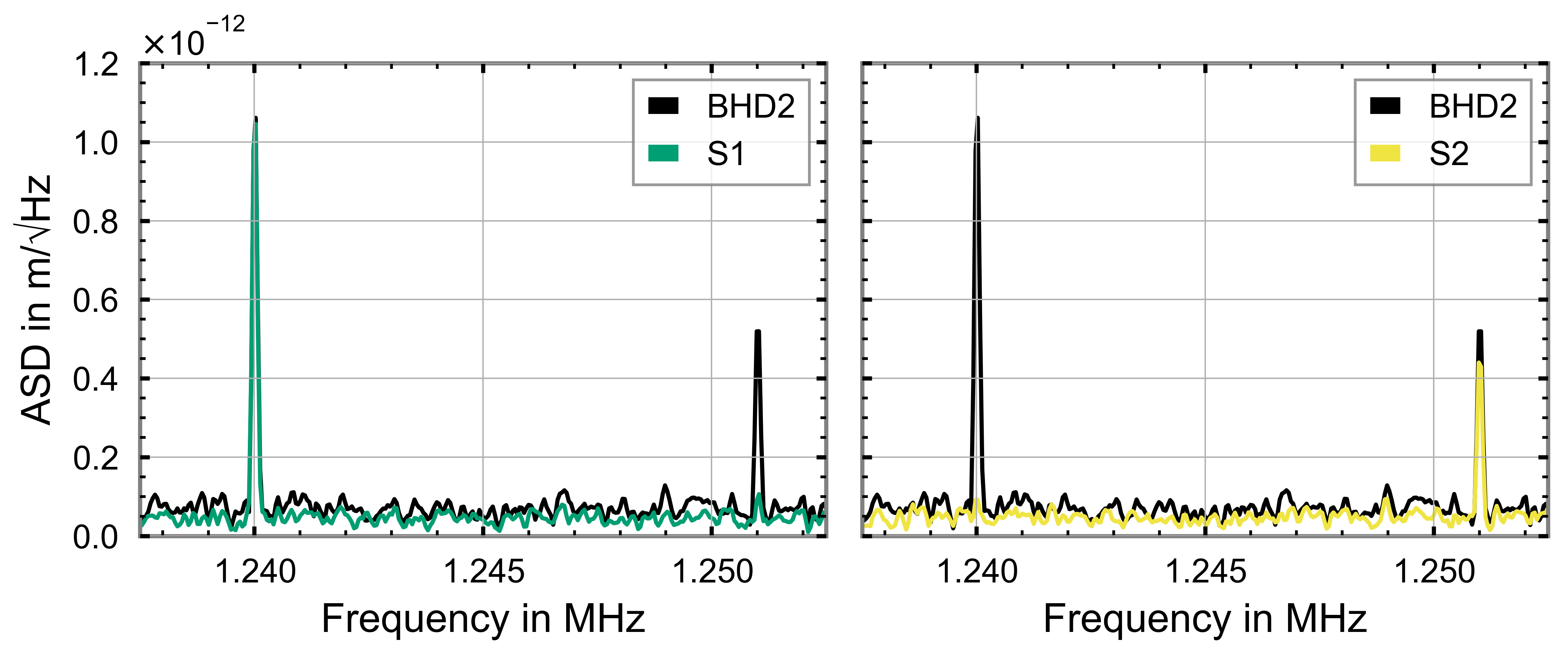}
            \caption{These amplitude spectral density plots show the scattered light noise suppression/the arm signal discrimination in the signal combinations $S_1$ and $S_2$. Left: $S_1$ shows a scattered light noise peak at \SI{1.251}{\mega\hertz} that is suppressed almost completely in comparison with the quadrature readout of BHD2 at the antisymmetric port, while the signal at \SI{1.24}{\mega\hertz} remains. Right: $S_2$ depicts a vanishing signal at \SI{1.24}{\mega\hertz}, while the scattered light noise peak largely remains.}
            \label{fig:suppression_wide}
        \end{figure*}

        Nevertheless, it is possible to demonstrate scattered light signal suppression and the individual readout of both interferometer arms. The combined signals $S_1$ and $S_2$ were computed according to \cref{eq:combined_signals_1} and their respective ASD's are depicted in comparison with the one of the BHD2 readout in \cref{fig:suppression_wide}. For the case of $S_1$ it is apparent that the scattered light noise peak is reduced by about \SI{13.2}{\decibel} in comparison to BHD2. Signal $S_2$ shows another, opposite result. Here, the signal peak is strongly suppressed, by about \SI{18.3}{\decibel}. Keep in mind that all signals are normalized to the signal peak, which has an RMS amplitude of about \SI{11.2}{\pico\meter}.

        \begin{figure}
            \centering
            \includegraphics[width=0.48\textwidth]{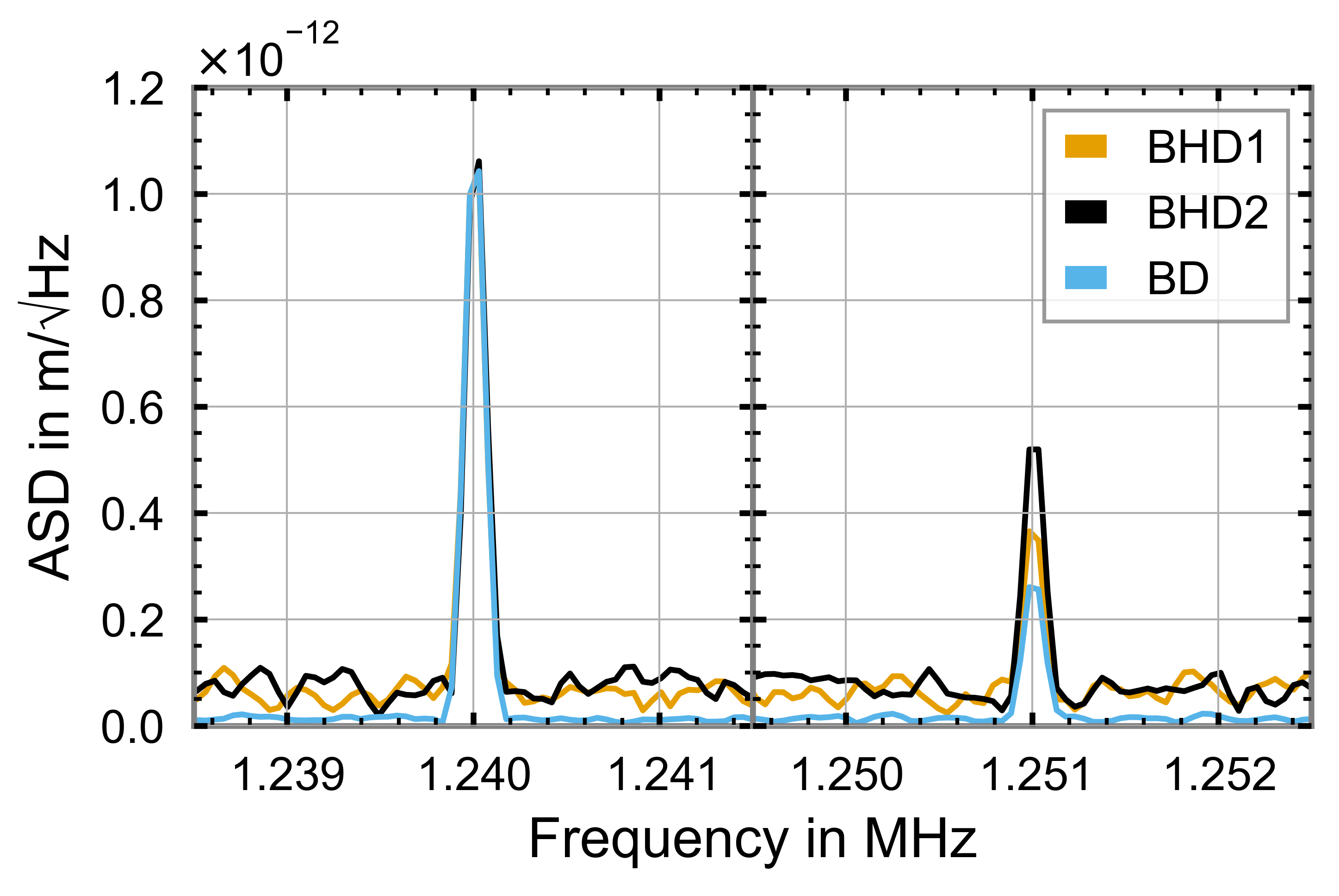}
            \caption{This amplitude spectral density plot shows the balanced homodyne readout signals (BHD1 and BHD2) in comparison with the balanced readout (BD). The balanced homodyne signals are normalized to the signal peak at $f_{sig} = \SI{1.24}{\mega\hertz}$, on the left side. The scatter peak is shown on the right side, at $f_{sc} = \SI{1.251}{\mega\hertz}$.}
            \label{fig:bhd_signals}
        \end{figure}
        
    \subsection{Discussion}
    \label{sec:exp_discussion}
        In the experiment described here, we were able to demonstrate scattered light suppression via implementation of one balanced homodyne detector at each interferometer port. This technique allows for the partial subtraction of noise (or signal in our experiment) stemming from either one of the interferometer arms by allowing for the discrimination between phase changes in either of the two arms. This stands in stark contrast to conventional readout techniques, in which a differential arm length change is measured.

        The measurement shows a suppression of scattered light induced signal by \SI{13.2}{\decibel} in one of the combined signals. Additionally, the suppression of the signal generated with a length modulation of one of the arms by \SI{18.3}{\decibel} in the other obtained signal was shown. Limiting here was most likely our implementation (discussed below), as there is no indication for a theoretical suppression limit in the theory laid out in \cref{sec:theory} except for the shot noise limit. This indicates that the demonstration of more suppression is possible, if the experiment can be improved. One could likely demonstrate a higher suppression by reducing electrical crosstalk in the experiment, by improving the modulation range of the scattered light peak to increase the signal strength or by a reduction of the readout noise, which we measured to be above the expected shot noise limit.

        Further, we assume that our experiment was limited by multiple aspects. On one hand, the locking scheme was limiting the precision of the quadrature readout. It was possible to lock one of the BHD's quadratures by using its readout as an error signal to control the local oscillator phase. Since the other BHD had to be adjusted to an orthogonal quadrature, this meant its signal was not usable directly as an error signal (the DC component was non-zero at the right quadrature). Thus, we adjusted it manually and the quadrature was consequently slightly detuned from being orthogonal. This could explain why the scattered light signal peaks inhibit different amplitudes, despite the coupling being equal in both quadratures, according to \cref{eq:combined_signals_1,eq:combined_signals_2}. Another possible explanation for our observation is the following. The readout at both ports was placed at differing distances from the central beam splitter. When setting up the experiment it was ensured this difference would be small enough such that the signal phase does not inherently change, i.e. below \SI{30}{\centi\meter}. It could, however, be possible that the overlap between the modes of the scattered light signal and the carrier signal differed from the symmetric and the antisymmetric port, which would not matter if there is no beam clipping, but we can not exclude this due to a small aperture at the Faraday isolator used to separate the symmetric output port. It is unclear, which effect dominates.

        For an operational implementation in a Michelson-based interferometer it will, however, be necessary to take a couple more measures. Note, that we did not discuss the choice of operating points, which could in practice be achieved by tuning both interferometer arms, such that signals in each arm show ideally in the quadrature readouts with respect to the limiting noise sources. 
        Evidently, here we only considered a simple Michelson interferometer, thus in the next section enhanced topologies are discussed for the application of scattered light noise subtraction in gravitational wave detectors. 

\section{Advanced interferometer topologies}
\label{sec:advanced}
    The demonstration of scattered light noise stemming from parasitic fields in the interferometer arms of a Michelson interferometer could be a crucial step towards better scattered light noise suppression in interferometric gravitational wave detectors.
    These interferometers operate at the dark port and employ more complex topologies, implementing arm cavities, power-recycling and signal-recycling to optimise shot-noise sensitivity and other technical noise reductions. 
    A comprehensive study of applying our technique to such interferometers goes beyond this letter, but we will discuss the role of each of these design elements shortly in the following. 
    To begin, the effect of arm cavities is considered. It can be observed that in \cref{eq:combined_signals_1,eq:combined_signals_2} for each obtained signal combination one respective interferometer arm does not contribute to the measured signal at all. This is independent of the actual phase, and a brief derivation can easily show that the suppression of the contribution of one arm is also irrespective of the actual field amplitude. Assuming a beam splitter in the same orientation as in \cref{fig:experiment} with electric fields entering from the X- and Y-direction, described by field amplitudes $\alpha_{X,Y} = Q_{X,Y} + iP_{X,Y}$. Then the outgoing amplitudes are
    \begin{equation}
        \begin{split}
            \alpha_{as} = \frac{1}{\sqrt{2}} (\alpha_X + i \alpha_Y),~
            \alpha_{s} = \frac{1}{\sqrt{2}} (\alpha_Y + i \alpha_X).
        \end{split}
    \end{equation}
    Consequently, the quadratures yield:
    \begin{equation}
        \begin{split}
            Q_{as} = \frac{1}{\sqrt{2}} (Q_X - P_Y),~
            P_{as} = \frac{1}{\sqrt{2}} (Q_Y + P_X).
        \end{split}
    \end{equation}
    and
    \begin{equation}
        \begin{split}
            Q_{s} = \frac{1}{\sqrt{2}} (Q_Y - P_X),~
            P_{s} =\frac{1}{\sqrt{2}} (Q_X + P_Y).
        \end{split}
    \end{equation}
    As above, combinations of orthogonal quadratures lead to a read out of separate quadratures of the fields in the interferometer arms. This shows, that the separation of signals is a general result. A quadrature readout at both exit ports of a beam splitter allows the reconstruction of two interfering fields entering the beam splitter. In the case of a gravitational wave detector, this enables the individual arm readout also for a Fabry-Perot Michelson interferometer. As such, this technique could be tested in interferometers that do not employ optical recycling techniques, such as gravitational wave detector prototype facilities \cite{westphal2012design}.

    However, it is assumed that power- and signal-recycling, as well as the necessary and desired dark-port operation, pose a challenge for the scheme. Remaining for a moment in the picture of a separation of the fields in the interferometer arms, the effect of, e.g., power-recycling would be to mix those amplitudes further before enabling the readout, as a large fraction of the field propagating from the interferometer towards the laser is again back-reflected into the interferometer to interfere again. This non-trivial process seems to make a separation of fields challenging.
 
    Another effect of a power-recycling scheme is that the removed amplitude noise is present in the field reflected from the power-recycling mirror towards the laser and might thus hinder a readout of that field. Consequently a different amplitude stabilization may be needed to implement the dual balanced readout scheme if using the field in reflection of the power-recycling cavity. Alternatively, one could look for a way to combine both schemes.
    
    Despite all the caveats presented here, making use of the presented dual balanced readout scheme seem almost futile, this study illustrates that additional detectors may improve our knowledge about an observed signal. This is stressed by the notable demonstrated use of auxiliary ports for the readout and removal of scattered light noise originating from the optical benches in transmission from the end test masses in Advanced Virgo\cite{Acernese2022,Ws2021}. Subsequent research could thus focus on examining each quadrature of each potential pick-off port -- i.e. transmission through test masses, other pick-offs inside the interferometer -- for potential use as a scattered light noise readout. 

    Coming back to the dual balanced readout scheme, there are several aspects that need to be studied in the future. Regarding the topology, it should be considered how to modify the readout to measure the scattered light noise, by moving the readouts or adding additional ones. Further, the effect of recycling techniques and resonant sideband extraction needs to be addressed. Apart from the interferometer topology it will be necessary to study the opto-mechanical coupling between stray light fields and the test masses and to consider the mitigation of the consequent additional scattering-induced radiation pressure noise. 
    The potentially necessary detection of quadratures from bright fields could, in the future, be realized by applying additional squeezed light when attenuating the bright fields \cite{vahlbruch_laser_2018} or by using opto-mechanical detection of bright fields \cite{TradNery2021}. Generally, implementing balanced homodyne detection at the symmetric port, even without including the just mentioned techniques, brings additional complexities. Balanced detection for the asymmetric port requires a number of additional optics and beam paths to use e.g. a pick-off from the central beam splitter as local oscillator \cite{fritschel_balanced_2014}. This effort would need to be effectively doubled to also enable balanced homodyne detection at the symmetric port.     
    Finally, all relevant noise couplings will have to be studied to find suitable combinations, either for detector operation or measurements of scattered light during commissioning.

\section{Conclusions}
\label{sec:discussion}
    In this letter a dual balanced homodyne detection scheme is reported that allows to read out arm length modulations in the two arms of a Michelson interferometer separately. The scheme consists of a dual balanced homodyne detector at both the symmetric port and the antisymmetric port and orthogonal quadratures are read out at each. Addition and subtraction of the measured signals allows to read out the electric field returning from each arm individually. A mathematical description of the scheme is provided in \cref{sec:theory}, while \cref{sec:experiment} describes the experimental setup and results of the demonstration. The separate readout can be used for the subtraction of scattered light noise, which in our experiment was deliberately injected into one of the interferometer arms using a partially reflective window and modulating the length of the stray light path. A scattered light signal suppression of \SI{13.2}{\decibel} is demonstrated here, while being able to read out a length modulation of the interferometer arm that was unaffected from scattering.
    It appears that the suppression of the scattered light noise contribution is limited by fundamental noise sources like shot noise, limiting technical noise sources and practical limitations in the experiment.

    This scheme is interesting for gravitational wave detectors -- which are typically based on the Michelson interferometer -- as scattered light noise poses a limitation of their sensitivity, especially at low frequencies. As discussed in \cref{sec:advanced}, it is expected that arm cavities do not affect the dual balanced homodyne detection scheme. Future efforts will, however, need to explore the effects of power- and signal-recycling on the readout and determine whether the scheme can be made compatible with these techniques. Further effort should also be directed towards addressing open questions, such as the limitation of the scheme through different types of laser noise, as well as if the readout will also be able to subtract the effect of opto-mechanical coupling of scattered light noise. Further, it is important to note that the sensitivity for each signal depends on the operating point. Thus for an optimal sensitivity it may be necessary to read out two orthogonal quadratures at each port or to find a compromise between the sensitivities of arm length changes in the two individual interferometer arms. Finally, the compatibility with quantum noise reduction by e.g. squeezed light needs to be investigated. 

    In summary, scattered light noise subtraction was demonstrated in a Michelson interferometer using combinations of signals from two balanced homodyne detectors. While this is a highly interesting technique for the use in gravitational wave detectors, a variety of questions need to be addressed before an implementation can be targeted, as the scheme can not be readily adopted in topologies more complicated than the simple case studied here.

\begin{acknowledgments}
    This research was funded by the Deutsche Forschungsgemeinschaft (DFG, German Research Foundation) under Germany's Excellence Strategy---EXC 2121 ``Quantum Universe''---390833306. The authors thank Leonie Eggers for the assistance during her time in the group. Further, the authors acknowledge valuable input from Michael Was and Benno Willke via the LIGO P\&P review.
\end{acknowledgments}


\begin{thebibliography}{33}
	\providecommand{\natexlab}[1]{#1}
	\providecommand{\url}[1]{\texttt{#1}}
	\expandafter\ifx\csname urlstyle\endcsname\relax
	\providecommand{\doi}[1]{doi: #1}\else
	\providecommand{\doi}{doi: \begingroup \urlstyle{rm}\Url}\fi
	
	\bibitem[Collaboration(2015{\natexlab{a}})]{LIGOScientific}
	LIGO~Scientific Collaboration.
	\newblock {Advanced LIGO}.
	\newblock \emph{Class. Quant. Grav.}, 32:\penalty0 074001, 2015{\natexlab{a}}.
	\newblock \doi{10.1088/0264-9381/32/7/074001}.
	
	\bibitem[Collaboration(2015{\natexlab{b}})]{VIRGO}
	VIRGO~Scientific Collaboration.
	\newblock {Advanced Virgo: a second-generation interferometric gravitational
		wave detector}.
	\newblock \emph{Class. Quant. Grav.}, 32\penalty0 (2):\penalty0 024001,
	2015{\natexlab{b}}.
	\newblock \doi{10.1088/0264-9381/32/2/024001}.
	
	\bibitem[Akutsu et~al.(2020)Akutsu, Ando, Arai, Arai, Araki, Araya, Aritomi,
	Aso, Bae, Bae, et~al.]{KAGRA}
	T~Akutsu, M~Ando, K~Arai, Y~Arai, S~Araki, A~Araya, N~Aritomi, Y~Aso, S~Bae,
	Y~Bae, et~al.
	\newblock {Overview of KAGRA: Detector design and construction history}.
	\newblock \emph{Progress of Theoretical and Experimental Physics},
	2021\penalty0 (5):\penalty0 05A101, 08 2020.
	\newblock ISSN 2050-3911.
	\newblock \doi{10.1093/ptep/ptaa125}.
	\newblock URL \url{https://doi.org/10.1093/ptep/ptaa125}.
	
	\bibitem[Lück et~al.(2010)Lück, Affeldt, Degallaix, Freise, Grote, Hewitson,
	Hild, Leong, Prijatelj, Strain, et~al.]{Luck:2010}
	H~Lück, C~Affeldt, J~Degallaix, A~Freise, H~Grote, M~Hewitson, S~Hild,
	J~Leong, M~Prijatelj, K~A Strain, et~al.
	\newblock {The upgrade of GEO600}.
	\newblock \emph{J. Phys. Conf. Ser.}, 228:\penalty0 012012, 2010.
	\newblock \doi{10.1088/1742-6596/228/1/012012}.
	
	\bibitem[Punturo et~al.(2010)Punturo, Abernathy, Acernese, Allen, Andersson,
	Arun, Barone, Barr, Barsuglia, Beker, et~al.]{ET}
	M~Punturo, M~Abernathy, F~Acernese, B~Allen, Nils Andersson, K~Arun, F~Barone,
	B~Barr, M~Barsuglia, M~Beker, et~al.
	\newblock The {{Einstein Telescope}}: A third-generation gravitational wave
	observatory.
	\newblock \emph{Classical and Quantum Gravity}, 27\penalty0 (19):\penalty0
	194002, 2010.
	
	\bibitem[Reitze et~al.(2019)Reitze, Adhikari, Ballmer, Barish, Barsotti,
	Billingsley, Brown, Chen, Coyne, Eisenstein, et~al.]{CosmicExplorer}
	David Reitze, Rana~X Adhikari, Stefan Ballmer, Barry Barish, Lisa Barsotti,
	GariLynn Billingsley, Duncan~A. Brown, Yanbei Chen, Dennis Coyne, Robert
	Eisenstein, et~al.
	\newblock {Cosmic Explorer: The U.S. Contribution to Gravitational-Wave
		Astronomy beyond LIGO}.
	\newblock 2019.
	\newblock URL \url{https://arxiv.org/abs/1907.04833}.
	
	\bibitem[Hall et~al.(2021)Hall, Kuns, Smith, Bai, Wipf, Biscans, Adhikari,
	Arai, Ballmer, Barsotti, et~al.]{CosmicExplorer2}
	Evan~D. Hall, Kevin Kuns, Joshua~R. Smith, Yuntao Bai, Christopher Wipf,
	Sebastien Biscans, Rana~X Adhikari, Koji Arai, Stefan Ballmer, Lisa Barsotti,
	et~al.
	\newblock Gravitational-wave physics with {Cosmic} {Explorer}: {Limits} to
	low-frequency sensitivity.
	\newblock \emph{Phys. Rev. D}, 103\penalty0 (12):\penalty0 122004, June 2021.
	\newblock \doi{10.1103/PhysRevD.103.122004}.
	\newblock URL \url{https://link.aps.org/doi/10.1103/PhysRevD.103.122004}.
	
	\bibitem[Billing et~al.(1983)Billing, Winkler, Schilling, R{\"u}diger,
	Maischberger, and Schnupp]{billing1983}
	H.~Billing, W.~Winkler, R.~Schilling, A.~R{\"u}diger, K.~Maischberger and
	L.~Schnupp.
	\newblock The {{Munich Gravitational Wave Detector Using Laser
			Interferometry}}.
	\newblock In Pierre Meystre and Marlan~O. Scully, editors, \emph{Quantum
		{{Optics}}, {{Experimental Gravity}}, and {{Measurement Theory}}}, pages
	525--566. Springer US, Boston, MA, 1983.
	\newblock ISBN 978-1-4613-3712-6.
	\newblock \doi{10.1007/978-1-4613-3712-6_23}.
	
	\bibitem[Vinet et~al.(1996)Vinet, Brisson, and Braccini]{vinet1996}
	Jean-Yves Vinet, Violette Brisson and Stefano Braccini.
	\newblock Scattered light noise in gravitational wave interferometric
	detectors: {{Coherent}} effects.
	\newblock \emph{Physical Review D}, 54\penalty0 (2):\penalty0 1276--1286, July
	1996.
	\newblock \doi{10.1103/PhysRevD.54.1276}.
	
	\bibitem[Vinet et~al.(1997)Vinet, Brisson, Braccini, Ferrante, Pinard, Bondu,
	and Tourni{\'e}]{vinet1997a}
	Jean-Yves Vinet, Violette Brisson, Stefano Braccini, Isidoro Ferrante, Laurent
	Pinard, Fran{\c c}ois Bondu and Eric Tourni{\'e}.
	\newblock Scattered light noise in gravitational wave interferometric
	detectors: {{A}} statistical approach.
	\newblock \emph{Physical Review D}, 56\penalty0 (10):\penalty0 6085--6095,
	November 1997.
	\newblock \doi{10.1103/PhysRevD.56.6085}.
	
	\bibitem[Schilling et~al.(1981)Schilling, Schnupp, Winkler, Billing,
	Maischberger, and Rudiger]{schilling_method_1981}
	R~Schilling, L~Schnupp, W~Winkler, H~Billing, K~Maischberger and A~Rudiger.
	\newblock A method to blot out scattered light effects and its application to a
	gravitational wave detector.
	\newblock \emph{Journal of Physics E: Scientific Instruments}, 14\penalty0
	(1):\penalty0 65--70, January 1981.
	\newblock ISSN 0022-3735.
	\newblock \doi{10.1088/0022-3735/14/1/018}.
	\newblock URL
	\url{http://stacks.iop.org/0022-3735/14/i=1/a=018?key=crossref.7c9f7c469a7612f1650865355822d8a3}.
	
	\bibitem[Schnupp et~al.(1985)Schnupp, Winkler, Maischberger, Rudiger, and
	Schilling]{schnupp_reduction_1985}
	L~Schnupp, W~Winkler, K~Maischberger, A~Rudiger and R~Schilling.
	\newblock Reduction of noise due to scattered light in gravitational wave
	antennae by modulating the phase of the laser light.
	\newblock \emph{Journal of Physics E: Scientific Instruments}, 18\penalty0
	(6):\penalty0 482--485, June 1985.
	\newblock ISSN 0022-3735.
	\newblock \doi{10.1088/0022-3735/18/6/005}.
	\newblock URL
	\url{http://stacks.iop.org/0022-3735/18/i=6/a=005?key=crossref.744d49a06a1bd9967d4f304bba5a1b0f}.
	
	\bibitem[Martynov et~al.(2016)Martynov, Hall, Abbott, Abbott, Abbott, Adams,
	Adhikari, Anderson, Anderson, Arai, et~al.]{martynov_2016}
	D.~V. Martynov, E.~D. Hall, B.~P. Abbott, R.~Abbott, T.~D. Abbott, C.~Adams,
	R.~X. Adhikari, R.~A. Anderson, S.~B. Anderson, K.~Arai, et~al.
	\newblock Sensitivity of the advanced ligo detectors at the beginning of
	gravitational wave astronomy.
	\newblock \emph{Phys. Rev. D}, 93:\penalty0 112004, Jun 2016.
	\newblock \doi{10.1103/PhysRevD.93.112004}.
	\newblock URL \url{https://link.aps.org/doi/10.1103/PhysRevD.93.112004}.
	
	\bibitem[Canuel et~al.(2013)Canuel, Genin, Vajente, and Marque]{Canuel2013}
	B.~Canuel, E.~Genin, G.~Vajente and J.~Marque.
	\newblock Displacement noise from back scattering and specular reflection of
	input optics in advanced gravitational wave detectors.
	\newblock \emph{Optics Express}, 21\penalty0 (9):\penalty0 10546, April 2013.
	\newblock \doi{10.1364/oe.21.010546}.
	\newblock URL \url{https://doi.org/10.1364/oe.21.010546}.
	
	\bibitem[Soni et~al.(2021)Soni, Austin, Effler, Schofield, Gonz{\'{a}}lez,
	Frolov, Driggers, Pele, Urban, Valdes, et~al.]{Soni2021}
	S~Soni, C~Austin, A~Effler, R~M~S Schofield, G~Gonz{\'{a}}lez, V~V Frolov, J~C
	Driggers, A~Pele, A~L Urban, G~Valdes, et~al.
	\newblock Reducing scattered light in {LIGO}'s third observing run.
	\newblock \emph{Classical and Quantum Gravity}, 38\penalty0 (2):\penalty0
	025016, January 2021.
	\newblock \doi{10.1088/1361-6382/abc906}.
	\newblock URL \url{https://doi.org/10.1088/1361-6382/abc906}.
	
	\bibitem[Nguyen et~al.(2021)Nguyen, Schofield, Effler, Austin, Adya, Ball,
	Banagiri, Banowetz, Billman, Blair, et~al.]{Nguyen2021}
	P~Nguyen, R~M~S Schofield, A~Effler, C~Austin, V~Adya, M~Ball, S~Banagiri,
	K~Banowetz, C~Billman, C~D Blair, et~al.
	\newblock Environmental noise in advanced {LIGO} detectors.
	\newblock \emph{Classical and Quantum Gravity}, 38\penalty0 (14):\penalty0
	145001, June 2021.
	\newblock \doi{10.1088/1361-6382/ac011a}.
	\newblock URL \url{https://doi.org/10.1088/1361-6382/ac011a}.

    \bibitem[Soni et~al.(2024)Soni, Glanzer, Effler, Frolov, González, Pele, and
    Schofield]{Soni2024}
    Siddharth Soni, Jane Glanzer, Anamaria Effler, Valera Frolov, Gabriela González, Arnaud Pele, and Robert Schofield.
    \newblock Modeling and reduction of high frequency scatter noise at LIGO Livingston.
    \newblock \emph{Classical and Quantum Gravity}, 41\penalty0 (13):\penalty0 135015, June 2024.
    \newblock ISSN 1361-6382.
    \newblock \doi{10.1088/1361-6382/ad494a}.
    \newblock URL \url{http://dx.doi.org/10.1088/1361-6382/ad494a}.
 
	\bibitem[Ottaway et~al.(2012)Ottaway, Fritschel, and Waldman]{Ottaway2012-vh}
	David~J Ottaway, Peter Fritschel and Samuel~J Waldman.
	\newblock Impact of upconverted scattered light on advanced interferometric
	gravitational wave detectors.
	\newblock \emph{Opt. Express}, 20\penalty0 (8):\penalty0 8329--8336, April
	2012.
	
	\bibitem[Chua et~al.(2014)Chua, Dwyer, Barsotti, Sigg, Schofield, Frolov,
	Kawabe, Evans, Meadors, Factourovich, et~al.]{Chua_2014}
	S~S~Y Chua, S~Dwyer, L~Barsotti, D~Sigg, R~M~S Schofield, V~V Frolov, K~Kawabe,
	M~Evans, G~D Meadors, M~Factourovich, et~al.
	\newblock Impact of backscattered light in a squeezing-enhanced interferometric
	gravitational-wave detector.
	\newblock \emph{Classical and Quantum Gravity}, 31\penalty0 (3):\penalty0
	035017, jan 2014.
	\newblock \doi{10.1088/0264-9381/31/3/035017}.
	\newblock URL \url{https://dx.doi.org/10.1088/0264-9381/31/3/035017}.
	
	\bibitem[Steinlechner et~al.(2013)Steinlechner, Bauchrowitz, Meinders,
	Müller-Ebhardt, Danzmann, and Schnabel]{steinlechner_quantum-dense_2013}
	Sebastian Steinlechner, Jöran Bauchrowitz, Melanie Meinders, Helge
	Müller-Ebhardt, Karsten Danzmann and Roman Schnabel.
	\newblock Quantum-dense metrology.
	\newblock \emph{Nature Photonics}, 7\penalty0 (8):\penalty0 626--630, August
	2013.
	\newblock ISSN 1749-4885, 1749-4893.
	\newblock \doi{10.1038/nphoton.2013.150}.
	\newblock URL \url{http://www.nature.com/articles/nphoton.2013.150}.
	
	\bibitem[Meinders and Schnabel(2015)]{Meinders2015}
	Melanie Meinders and Roman Schnabel.
	\newblock Sensitivity improvement of a laser interferometer limited by
	inelastic back-scattering, employing dual readout.
	\newblock \emph{Classical and Quantum Gravity}, 32\penalty0 (19):\penalty0
	195004, September 2015.
	\newblock \doi{10.1088/0264-9381/32/19/195004}.
	\newblock URL \url{https://doi.org/10.1088/0264-9381/32/19/195004}.
	
	\bibitem[Ast et~al.(2016)Ast, Steinlechner, and Schnabel]{ast2016}
	Melanie Ast, Sebastian Steinlechner and Roman Schnabel.
	\newblock Reduction of {Classical} {Measurement} {Noise} via {Quantum}-{Dense}
	{Metrology}.
	\newblock \emph{Physical Review Letters}, 117\penalty0 (18):\penalty0 180801,
	October 2016.
	\newblock \doi{10.1103/PhysRevLett.117.180801}.
	\newblock URL \url{https://link.aps.org/doi/10.1103/PhysRevLett.117.180801}.
	\newblock tex.ids: ast2016a.
	
	\bibitem[Ast(2017)]{ast_quantum-dense_2017}
	Melanie Ast.
	\newblock \emph{Quantum-dense metrology for substraction of back-scatter
		disturbances in gravitational-wave detection}.
	\newblock {PhD} {Thesis}, Dissertation, Gottfried Wilhelm Leibniz Universität
	Hannover, 2017.
	
	\bibitem[Fleddermann et~al.(2018)Fleddermann, Diekmann, Steier, Tr\"{o}bs,
	Heinzel, and Danzmann]{Fleddermann2018}
	Roland Fleddermann, Christian Diekmann, Frank Steier, Michael Tr\"{o}bs,
	Gerhard Heinzel and Karsten Danzmann.
	\newblock Sub-pm/$\sqrt{\mathrm{hz}}^{-1}$ non-reciprocal noise in the {LISA}
	backlink fiber.
	\newblock \emph{Classical and Quantum Gravity}, 35\penalty0 (7):\penalty0
	075007, February 2018.
	\newblock \doi{10.1088/1361-6382/aaa276}.
	\newblock URL \url{https://doi.org/10.1088/1361-6382/aaa276}.
	
	\bibitem[Steier(2008)]{steier2008a}
	Frank Steier.
	\newblock \emph{Interferometry Techniques for Spaceborne Gravitational Wave
		Detectors}.
	\newblock PhD thesis, Hannover, Univ., Diss., 2008.
	
	\bibitem[Fox(2006)]{Fox2006-yd}
	Mark Fox.
	\newblock \emph{Quantum Optics}.
	\newblock Oxford Master Series in Physics. Oxford University Press, London,
	England, April 2006.
	
	\bibitem[Welch(1967)]{Welch1967}
	P.~Welch.
	\newblock The use of fast fourier transform for the estimation of power
	spectra: A method based on time averaging over short, modified periodograms.
	\newblock \emph{{IEEE} Transactions on Audio and Electroacoustics}, 15\penalty0
	(2):\penalty0 70--73, June 1967.
	\newblock \doi{10.1109/tau.1967.1161901}.
	\newblock URL \url{https://doi.org/10.1109/tau.1967.1161901}.
	
	\bibitem[Westphal et~al.(2012)Westphal, Bergmann, Bertolini, Born, Chen,
	Cumming, Cunningham, Dahl, Gr{\"a}f, Hammond, et~al.]{westphal2012design}
	T~Westphal, G~Bergmann, A~Bertolini, M~Born, Y~Chen, AV~Cumming, L~Cunningham,
	K~Dahl, C~Gr{\"a}f, G~Hammond, et~al.
	\newblock Design of the 10 m aei prototype facility for interferometry studies:
	A brief overview.
	\newblock \emph{Applied Physics B}, 106:\penalty0 551--557, 2012.
	
	\bibitem[Acernese et~al.(2022)Acernese, Agathos, Ain, Albanesi, Allocca, Amato,
	Andrade, Andres, Andrić, Ansoldi, et~al.]{Acernese2022}
	F~Acernese, M~Agathos, A~Ain, S~Albanesi, A~Allocca, A~Amato, T~Andrade,
	N~Andres, T~Andrić, S~Ansoldi, et~al.
	\newblock Calibration of advanced virgo and reconstruction of the detector
	strain h(t) during the observing run o3.
	\newblock \emph{Classical and Quantum Gravity}, 39\penalty0 (4):\penalty0
	045006, January 2022.
	\newblock ISSN 1361-6382.
	\newblock \doi{10.1088/1361-6382/ac3c8e}.
	\newblock URL \url{http://dx.doi.org/10.1088/1361-6382/ac3c8e}.
	
	\bibitem[Wąs et~al.(2021)Wąs, Gouaty, and Bonnand]{Ws2021}
	M~Wąs, R~Gouaty and R~Bonnand.
	\newblock End benches scattered light modelling and subtraction in advanced
	virgo.
	\newblock \emph{Classical and Quantum Gravity}, 38\penalty0 (7):\penalty0
	075020, March 2021.
	\newblock ISSN 1361-6382.
	\newblock \doi{10.1088/1361-6382/abe759}.
	\newblock URL \url{http://dx.doi.org/10.1088/1361-6382/abe759}.
	
	\bibitem[Vahlbruch et~al.(2018)Vahlbruch, Wilken, Mehmet, and
	Willke]{vahlbruch_laser_2018}
	Henning Vahlbruch, Dennis Wilken, Moritz Mehmet and Benno Willke.
	\newblock Laser {Power} {Stabilization} beyond the {Shot} {Noise} {Limit}
	{Using} {Squeezed} {Light}.
	\newblock \emph{Physical Review Letters}, 121\penalty0 (17):\penalty0 173601,
	October 2018.
	\newblock \doi{10.1103/PhysRevLett.121.173601}.
	\newblock URL \url{https://link.aps.org/doi/10.1103/PhysRevLett.121.173601}.
	\newblock Publisher: American Physical Society.
	
	\bibitem[Nery et~al.(2021)Nery, Venneberg, Aggarwal, Cole, Corbitt, Cripe,
	Lanza, and Willke]{TradNery2021}
	Marina~Trad Nery, Jasper~R. Venneberg, Nancy Aggarwal, Garrett~D. Cole, Thomas
	Corbitt, Jonathan Cripe, Robert Lanza and Benno Willke.
	\newblock Laser power stabilization via radiation pressure.
	\newblock \emph{Opt. Lett.}, 46\penalty0 (8):\penalty0 1946--1949, Apr 2021.
	\newblock \doi{10.1364/OL.422614}.
	\newblock URL \url{https://opg.optica.org/ol/abstract.cfm?URI=ol-46-8-1946}.

    \bibitem[Fritschel et~al.(2014)Fritschel, Evans, and
	Frolov]{fritschel_balanced_2014}
	Peter Fritschel, Matthew Evans, and Valery Frolov.
	\newblock Balanced homodyne readout for quantum limited gravitational wave
	detectors.
	\newblock \emph{Optics Express}, 22\penalty0 (4):\penalty0 4224--4234, February
	2014.
	\newblock ISSN 1094-4087.
	\newblock \doi{10.1364/OE.22.004224}.
	\newblock URL \url{https://opg.optica.org/oe/abstract.cfm?uri=oe-22-4-4224}.
	\newblock Publisher: Optica Publishing Group.
    
\end{thebibliography}
\end{document}